\documentclass[a4paper,11pt]{article}

\usepackage{pos}
\graphicspath{{ICRC2023-BSM3/}{Figures/}}

\title{Search for the rare interactions of neutrinos from distant point sources with IceCube Neutrino Telescope}

\ShortTitle{Neutrino Rare Interaction Search with IceCube}

\author{The IceCube Collaboration \\{\normalsize \normalfont(a complete list of authors can be found at the end of the proceedings)}\\}

\emailAdd{woosik.kang@icecube.wisc.edu}
\emailAdd{rott@physics.utah.edu}

\abstract{

The recent discovery and evidence of neutrino signals from distant sources, \emph{TXS 0506+056} and \emph{NGC 1068} respectively, provide opportunities to search for rare interactions of neutrinos that they might encounter on their paths. One potential scenario of interest is the interaction between neutrinos and dark matter, which is invisible and expected to be abundantly spread over the Universe. Various astrophysical observations have implied the existence of dark matter. When high-energy neutrinos from extragalactic sources interact with dark matter during their propagation, their spectra might show suppressions at specific energy ranges, where such interactions occur. These attenuation signatures from the interaction might be measurable on Earth with large neutrino telescopes such as the IceCube Neutrino Observatory. This analysis is focused on the search for rare interactions of high-energy neutrinos from the IceCube-identified astrophysical neutrino sources with dark matter in sub-GeV masses and several benchmark mediator cases using the upgoing track-like events. In this poster, sensitivity studies about the interaction of neutrinos and dark matter are presented.

\vspace{4mm}
{\bfseries Corresponding authors:}
Woosik Kang$^{1*}$, Carsten Rott$^{1,2}$\\
{$^{1}$ \itshape Department of Physics, Sungkyunkwan University, Suwon 16419, Rep. of Korea}\\
{$^{2}$ \itshape Department of Physics \& Astronomy, University of Utah, Salt Lake City, UT 84112, USA}\\
\\[4mm]
$^*$ Presenter

\ConferenceLogo{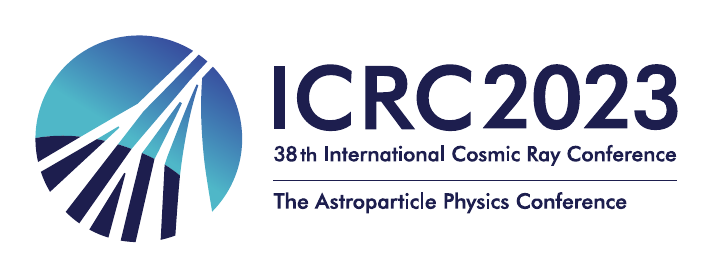}

\FullConference{The 38th International Cosmic Ray Conference (ICRC2023)\\ 26 July -- 3 August, 2023\\ Nagoya, Japan}
}

\begin{document}

\maketitle

\section{Astrophysical Neutrinos and IceCube}\label{sec1}
Neutrinos are regarded as excellent astrophysical messenger particles as their propagation is not disturbed by magnetic fields inducing deflections or attenuated from interactions with the baryonic matter due to their neutral electric charge and weak interactions. The observed astrophysical neutrinos can reveal the locations where they were produced in the Universe. This allows the high-energy astrophysical neutrinos to be used further to probe the interior of dense and extreme environments of the Universe, which are believed to produce the high-energy neutrinos, such as flaring blazars, galaxies with active galactic nuclei, etc.

The IceCube Neutrino Telescope~\cite{Aartsen:2016nxy} is the world's largest neutrino detector located at the geographic South Pole. The telescope utilises one cubic-kilometre volume of the ultra-pure Antarctic ice at depths between 1450~m and 2450~m as detector medium. This giga-ton scale in-ice detector is equipped with 5,160 Digital Optical Modules (DOMs), and each DOM instruments a single downward-facing 10-inch photomultiplier tube (PMT) within a pressure-resistant glass sphere. These PMTs can observe the Cherenkov radiation emitted by relativistic charged particles created by neutrino interactions in the detector medium.

Since the first result for the observation of extraterrestrial neutrinos in 2013~\cite{IceCube:2013low}, the observatory has identified multiple high-energy astrophysical neutrino sources as of now: the known gamma-ray blazar, \emph{TXS 0506+056} in 2018~\cite{IceCube:2018dnn, IceCube:2018cha}, the nearby active galaxy, \emph{NGC 1068} in 2022~\cite{IceCube:2022der}, and the Galactic Plane of the Milky Way galaxy in 2023~\cite{IceCube:2023gpp}.
Following the observation of high-energy astrophysical neutrinos, IceCube keeps accumulating larger and larger astrophysical neutrino data and source locations, which allows explorations for new physical phenomena including the effects of neutrino propagation from interactions occurring on their journeys through the cosmos.

\section{Neutrino Rare Interactions}\label{sec2}
The properties of neutrinos are less well understood than those of other fundamental particles in the Standard Model (SM) of particle physics. Nevertheless, it also means neutrinos can be used as a probe for physics Beyond the Standard Model (BSM) by connecting the known and unknown sectors. In particular, there could be new interactions of neutrinos that are introduced by BSM models motivated to resolve the issues and tensions in astrophysical observations and Cosmology~\cite{Cherry:2014xra,Barenboim:2019tux,Berryman:2022hds}. 
Given the cosmological baseline from sources producing the high-energy astrophysical neutrinos, the BSM interactions can result in significant effects as they are integrated for the travelling of astrophysical neutrinos. To draw conclusions for these rare neutrino interactions with the IceCube high-energy astrophysical neutrino data, there are diverse approaches proposed considering the models such as the neutrino self-interactions with cosmic neutrino backgrounds ($C \nu B$)~\cite{Bustamante:2020mep}, or the long-range flavoured interactions with cosmological electron repositories~\cite{Bustamante:2018mzu}. 

From the models where the new interaction cross-section scales with neutrino energy, the interactions in the present Universe can dissipate the high-energy astrophysical neutrinos in propagation and lead to the suppression of astrophysical neutrino flux at Earth that could be observed by neutrino telescopes such as IceCube. In this proceeding, the interactions between neutrino and Dark Matter (DM) introducing the distortion of high-energy astrophysical neutrino spectra at Earth are considered as benchmark test cases. The details of $\nu$-DM interactions are discussed in Section~\ref{subsec2.1}.

\subsection{Interaction between Neutrino and Dark Matter}\label{subsec2.1}
The existence and the nature of DM is one of the most pressing unsolved problems in physics today. The conventional indirect experimental efforts to search for signals from DM largely focus on the products of DM annihilation or DM decay. However, the existence of the DM annihilation process producing SM particles as its final products implies the existence of the elastic scattering process between DM and SM particles. 
Moreover, the scattering of DM and neutrinos, assuming light DM mass which can produce a large number density of DM particles all over the universe, increases the opportunities of finding the interaction. 

To search for the interaction of neutrinos and DM, complementary experimental approaches are proposed to use the large-statistic dataset observed by neutrino telescopes. One of the approaches utilised the diffuse astrophysical neutrino flux, which appears consistent with an isotropic distribution at the highest energies, by considering the cosmological DM and DM halo of the Milky Way galaxy that leads to larger flux attenuation at or nearby the Galactic Centre~\cite{Arguelles:2017atb}. The study used the public data from IceCube High Energy Starting Event (HESE) selection~\cite{IceCube:2015hes} to test several interaction models, and the same method was applied in a full IceCube analysis~\cite{IceCube:2022clp}. There is another approach proposed in~\cite{Choi:2019ixb} to use neutrinos from distant neutrino point sources, which showed competitive bounds can be obtained by using the public event and its source information from IC170922A~\cite{IceCube:2018dnn} with consideration of the flux attenuation from the cosmological DM and DM halo of the Milky Way galaxy for travelling of neutrinos to the Earth. This analysis is motivated by the study whereby the neutrino flux from a distant point source is expected to undergo interactions with DM in propagation over cosmological distance, which leads to flux suppression at the Earth.

The previous studies solely considered the cosmological DM and the Milky Way DM halo, but this analysis includes Dark Matter in the vicinity of an astrophysical neutrino source as well. The detailed description of DM contributions considered in this analysis can be found in Section~\ref{subsec3.1}. Furthermore, this analysis is designed to use the full IceCube data described in Section~\ref{subsec3.4} to study the spectra distortion from interactions with the statistical analysis method delineated in Section~\ref{subsec3.5}. 

\section{Analysis with a benchmark model}\label{sec3}
A statistical analysis method is used to perform the hypothesis test to compare the Null hypothesis and the BSM alternative. The Null hypothesis expects that there exists a point source, resulting in $n_{s}$ signal events in the IceCube's observed data, whose neutrino spectrum is consistent with standard astrophysical neutrino source expectations, such as single power-law, broken power-law, spectral cut-off, etc~\cite{IceCube:2021uhz}. However, the BSM alternative expects BSM effects to be superimposed on the standard astrophysical spectrum defined in the Null hypothesis. Therefore, both hypotheses agree to assume the excess of neutrino signal events over the background events at the location of a source in the sky, but they expect different shapes of the neutrino spectrum observed by IceCube.

As given in Section~\ref{sec2}, this analysis considers the $\nu$-DM interaction as the BSM hypothesis. Letting the other spectral shapes considered as part of systematic effects, the Null hypothesis employs a single power-law function to describe the neutrino flux with the energy of neutrinos $E_{\nu}$, the spectral index $\gamma$, and the flux normalisation at 100~GeV of neutrino energy $\Phi_{0}$:
\begin{equation}
    \label{eq:spl}
    \Phi = \Phi_{0} \left( \frac{E_{\nu}}{100 \ \mathrm{GeV}} \right)^{-\gamma}. 
\end{equation}

\subsection{Dark Matter contributions to the interaction}\label{subsec3.1}
The evolution of neutrino flux $\Phi$ from a source due to the interactions between neutrino and Dark Matter can be described by the Boltzmann equation with the total cross-section $\sigma_{\nu\chi}$ and the differential cross-section $\frac{d\sigma_{\nu\chi}}{dE_{\nu}}$ which are the functions of neutrino energy $E_{\nu}$:
\begin{equation}
    \label{eq:bolzmann_equation}
    \frac{d\Phi}{d\tau} = -\sigma_{\nu\chi}(E_{\nu})\Phi + \int_{E_{\nu}}^{\infty} dE'_{\nu} \frac{d\sigma_{\nu\chi}}{dE_{\nu}}(E'_{\nu} \rightarrow E_{\nu})\Phi,
\end{equation}
where $\tau$ is the DM column density that describes the amount of DM particles along the line of sight ($l.o.s$) to a source. In detail, the first term on the right-hand side of Equation~\ref{eq:bolzmann_equation} gives the general attenuation of the initial flux, and the second term describes the re-distribution of neutrino energies after the scattering. DM column density, $\tau$, is derived from $\tau=\Sigma_{DM}/m_{DM}$ which is the ratio between the mass of DM particle, $m_{DM}$, and the accumulated DM mass along l.o.s., $\Sigma_{DM}=\int_{l.o.s.}\rho_{\chi}(r)dr$. Here, DM mass density at a position $r$ is given as $\rho_{\chi}(r)$. 

Considering Dark Matter distributions along l.o.s to a source, the contributions to the DM column density can be separated into three components. One is the distribution of DM within our Milky Way galaxy that is described by the DM mass density profile $\rho_{galactic}(\mathbf{x})$. Another one is DM in the intergalactic free space with the density $\rho_{cosmological}(z)$ that can be calculated from the cosmological constants and the distance between the Earth and a source. And the last one is the DM in the vicinity of a source producing high-energy astrophysical neutrinos, which is profiled as $\rho_{source}(\mathrm{r})$. The total DM mass along l.o.s is the sum of three integrated densities described as 
\begin{equation}
    \label{eq:dm_contributions}
    \Sigma_{DM} = \int_{l.o.s.} \rho_{galactic}(\mathbf{x})dl + \int_{l.o.s.} \rho_{cosmological}(z)dl + \int_{l.o.s.} \rho_{source}(r)dl. 
\end{equation}
In this analysis, the standard Navarro-Frank-White (NFW) profile~\cite{Navarro:1996gj} is chosen to describe the contribution from the galactic DM halo~\cite{Nesti:2013uwa}, and the cosmological constants from Planck 2018 data~\cite{Planck:2018vyg} are used to calculate the cosmological DM contribution from the redshift of each source~\cite{Farzan:2014gza}. Also, the DM spike density profile with the standard NFW profile is adopted to estimate the DM distribution surrounding a source galaxy with a supermassive black hole (SMBH) as its core~\cite{Ferrer:2022kei}.

\subsection{Benchmark Model as signals}\label{subsec3.2}
\begin{figure}[t]
    \centering
    \includegraphics[width=0.5\linewidth]{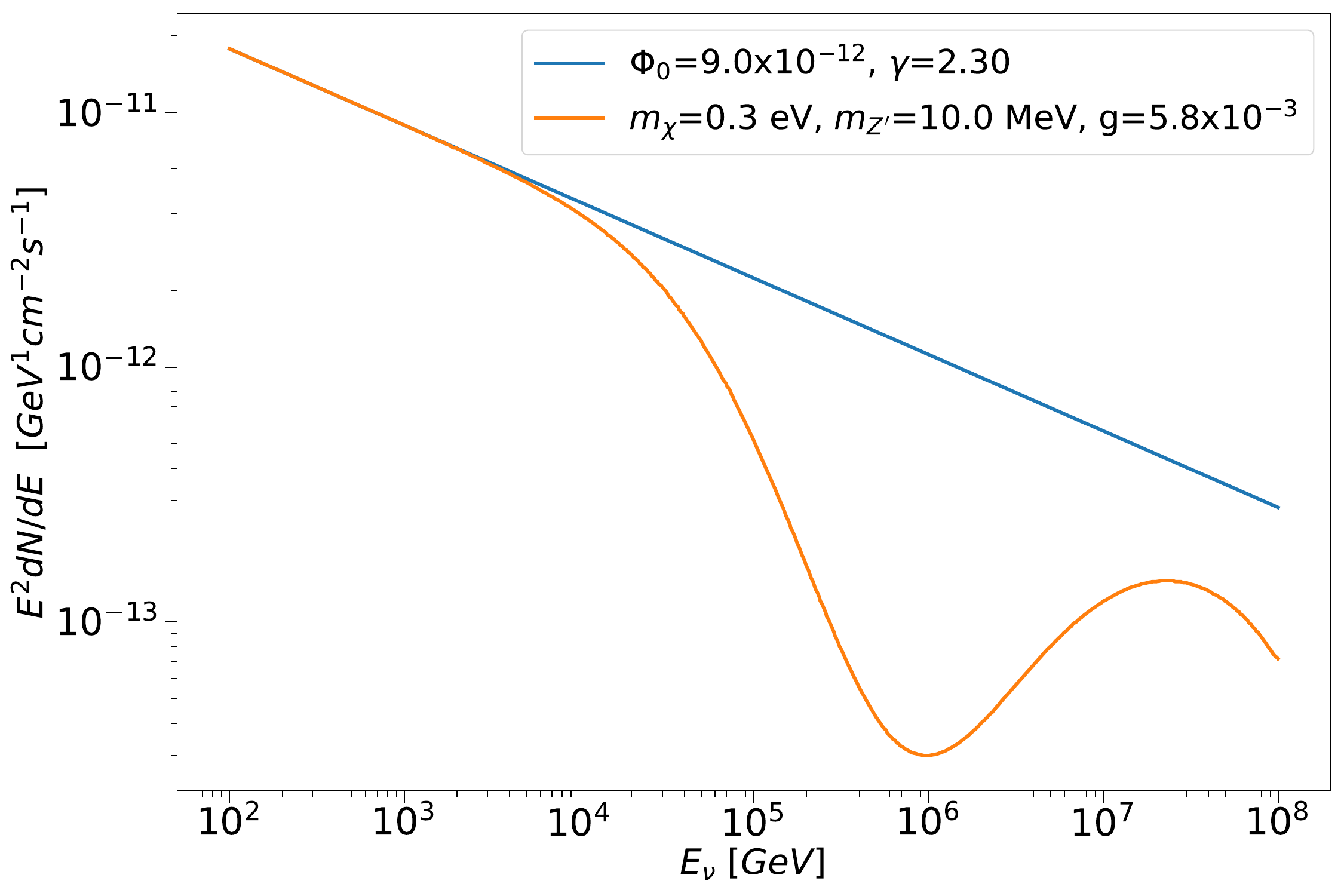}
    \caption{\textbf{Benchmark neutrino spectra from a point source at Earth} Benchmark spectra with and without the $\nu$-DM interactions are shown. The blue line is a single power-law spectrum with $\gamma=2.3$ and $\Phi_{0}=9 \times 10^{-12} \ \mathrm{GeV^{1}cm^{-2}s^{-1}}$ that corresponds to the Null hypothesis. The orange line is basically the same single power-law spectrum but has the 'dip' shape around $100 \ \mathrm{TeV}$ as an interaction signature the interaction with assumptions of $m_{\chi}=0.3 \ \mathrm{eV}$, $m_{Z'}=10 \ \mathrm{MeV}$, $g=5.8 \times 10^{-3}$ that corresponds to the BSM hypothesis.} 
    \label{fig:benchmark_spectra}
\end{figure}

To develop an analysis framework, two benchmark spectra are prepared as shown in Figure~\ref{fig:benchmark_spectra}; the blue line shows the Null hypothesis spectrum whereas the orange line displays the spectrum of the BSM alternative where the outstanding 'dip' feature exists around $100 \ \mathrm{TeV}$. 

For the BSM hypothesis, a model of the interaction between the scalar DM particles and neutrinos mediated via new vector particles ($Z'$) is assumed. The model was introduced in~\cite{Pandey:2018wvh} and an example diffuse spectrum from the model was presented in~\cite{Pandey:2020pgf}. This analysis converts the example spectrum to the BSM benchmark case in Figure~\ref{fig:benchmark_spectra} by restricting the solid angle to a point-like source in the sky. According to the presentation~\cite{Pandey:2020pgf}, the $\nu$-DM interaction whose cross-section scales with neutrino energy but approximately flattens at the highest energies creates the 'dip' signature from the evolution of flux. The initial spectral index of the flux and the flux normalisation at 100~GeV are given as $\gamma=2.3$ and $\Phi_{0}=9 \times 10^{-12} \ \mathrm{GeV^{1}cm^{-2}s^{-1}}$ respectively. The mass of the Dark Matter particle, the mass of the mediator particle, and the interaction coupling are set to be $m_{\chi}=0.3 \ \mathrm{eV}$, $m_{Z'}=10 \ \mathrm{MeV}$, $g=5.8 \times 10^{-3}$ respectively.

The Null hypothesis flux shares $\gamma$ and $\Phi_{0}$ with the BSM flux, but it is just extrapolated to higher energies with the single power-law in Equation~\ref{eq:spl}.
As expected, there is no interaction signature over all energies.

\subsection{Backgrounds}\label{subsec3.3}
Backgrounds for this analysis consist of atmospheric neutrinos and diffuse astrophysical neutrinos. Even though both conventional atmospheric neutrinos and prompt atmospheric neutrinos contribute to backgrounds, the total atmospheric neutrino flux is considered uniform in right ascension. Besides the astrophysical neutrino flux from a source, the flux of diffuse astrophysical neutrinos originating from the other region of the Universe away from the source location is expected to be isotropic and thus uniform in right ascension. Under these expectations, the right ascension of observed data is replaced with random numbers between $0^{\circ}$ and $360^{\circ}$ to estimate the background event distribution, called the event scrambling method. Contribution from atmospheric muons is considered negligible as the analysis focuses on the Northern sky (see Section~\ref{subsec3.4}).

\subsection{Data Sample}\label{subsec3.4}
Since all of the IceCube-identified astrophysical neutrino point sources are located in the Northern sky, and the analysis requires a good angular resolution to point back the direction of each neutrino event, the analysis considers the through-going track-like events from the Northern sky that mostly are induced by upward-going $\nu_{\mu}$ and $\bar{\nu}_{\mu}$ with respect to the detector. The data sample used in this analysis is a set of those neutrino events, selected by the well-established event selection also used in previous IceCube analyses~\cite{IceCube:2021uhz,IceCube:2022der}, from $\sim11.3$~years of observed data. The event selection achieves the median angular resolution to be $0.4^{\circ}$ at 100~TeV~\cite{IceCube:2022der}, which is preferred in the point source searches that are expected to produce high-energy astrophysical neutrinos. The sample contains events recorded in declinations ranging from $-5^{\circ}$ to $90^{\circ}$ and in energies above 100~GeV. Due to the absorption in the Earth or in the ice overburden, the atmospheric muon background is sufficiently suppressed which increases the purity of the data sample. 

\subsection{Statistical Methodology}\label{subsec3.5}
The previous IceCube point source searches~\cite{IceCube:2018ndw,IceCube:2019cia} utilised the unbinned likelihood method \cite{Braun:2008bg} to search for the astrophysical neutrino by maximising the signal-over-background likelihood estimation. This analysis employs a similar approach but with some modifications in the unbinned likelihood function $\mathcal{L}$ to include the interaction parameters as expressed in Equation~\ref{eq:likelihood}: 
\begin{equation}
    \label{eq:likelihood}
    \mathcal{L}(n_{s}) = \prod_{i=1}^{N} \left( \frac{n_{s}}{N}\mathcal{S}_{i}(\alpha_{i},\delta_{i},E_{i}|\gamma,\Phi_{0},m_{\chi},m_{\phi},g) + \left( 1- \frac{n_{s}}{N} \right)\mathcal{B}_{i}(\alpha_{i},\delta_{i},E_{i}|\gamma,\Phi_{0}) \right),
\end{equation}
where $\mathcal{S}$ and $\mathcal{B}$ are the probability density functions (PDFs) for signal and background described as 
\begin{equation}
    \label{eq:pdfs}
    \mathcal{S}_{i} = \underbrace{S(\alpha_{i}, \sin\delta_{i})}_\text{spatial PDF} \underbrace{S(E_{i},\sin\delta_{i})}_\text{energy PDF}\, , \ \mathcal{B}_{i} = \underbrace{B(\sin\delta_{i})}_\text{spatial PDF} \underbrace{B(E_{i},\sin\delta_{i})}_\text{energy PDF}.
\end{equation}
The PDFs are used to construct $\mathcal{L}$ for evaluating total $N$ observed events with observables; the reconstructed right ascension $\alpha$, the reconstructed declination $\delta$, and the reconstructed neutrino energy $E$ for $i$-th event.
Each $\mathcal{S}$ and $\mathcal{B}$ of $i$-th event are approximated as the product of a spatial PDF and an energy PDF~\cite{Braun:2008bg}. 
Each local hypothesis test gets its own $\mathcal{S}$ from dependencies on $m_{\chi}$, $m_{\phi}$, and $g$. $\mathcal{B}$ is obtained by data scrambling as described in Section~\ref{subsec3.4}. 
Each hypothesis has its own likelihood function; $\mathcal{L}_{BSM}$ for the BSM hypothesis, $\mathcal{L}_{Null}$ for the Null hypothesis, and $\mathcal{L}_{BG}$ for the Background-only hypothesis. The test statistic for this analysis ($\mathcal{TS}$) is defined as the difference of the log-likelihoods for BSM hypothesis and Null hypothesis, which becomes $\Delta TS$, described as  
\begin{align}
    \mathcal{TS} & = -2 \cdot sign(n_{s}) \cdot \log\left[ \frac{\mathcal{L}_{Null}}{\mathcal{L}_{BSM}}\right] \label{eq:test_statistics1} \\
    & = -2 \cdot sign(n_{s}) \cdot \left( \log\left[ \frac{\mathcal{L}_{BG}}{\mathcal{L}_{BSM}}\right] - \log\left[ \frac{\mathcal{L}_{BG}}{\mathcal{L}_{Null}}\right] \right) = TS_{BSM} - TS_{Null} \equiv \Delta TS. \label{eq:test_statistics2} 
\end{align}
The definition of the test statistic ($TS$) with the Background-only hypothesis in Equation~\ref{eq:test_statistics2} is technically identical to the $TS$ definition used in the previous IceCube point source analyses~\cite{IceCube:2018ndw,IceCube:2019cia}.

$\mathcal{L}$ for both Null and BSM hypotheses are maximised by fitting for the expected number of signal neutrino events, $n_{s}$, with the information of $N$ events. $\mathcal{L}_{BG}$ only takes $n_{s}=0$ to give no signal event, and $\mathcal{L}_{Null}$ only takes $g_{\nu\chi}=0$ to have no interaction signature in the spectrum at the Earth. $TS$ is allowed to be both positive and negative values by accepting the sign of $n_{s}$, which account for under- and over-fluctuation of background events at the source location in the sky respectively. 

\section{Analysis Status}\label{sec4}
\begin{figure}[t]
    \centering
    \includegraphics[width=0.329\linewidth]{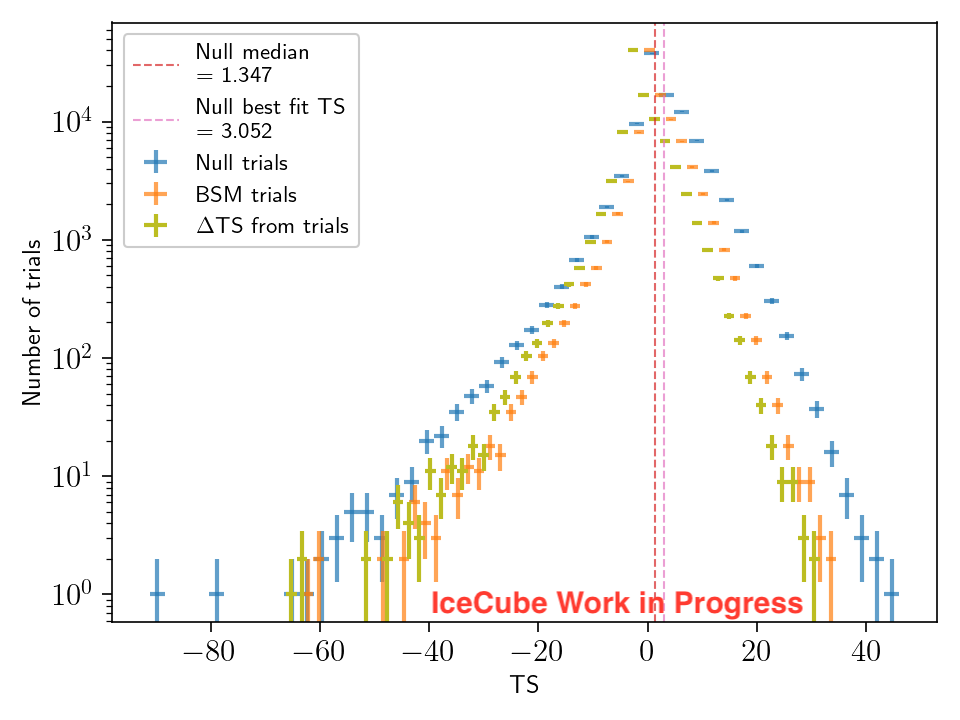}
    \includegraphics[width=0.329\linewidth]{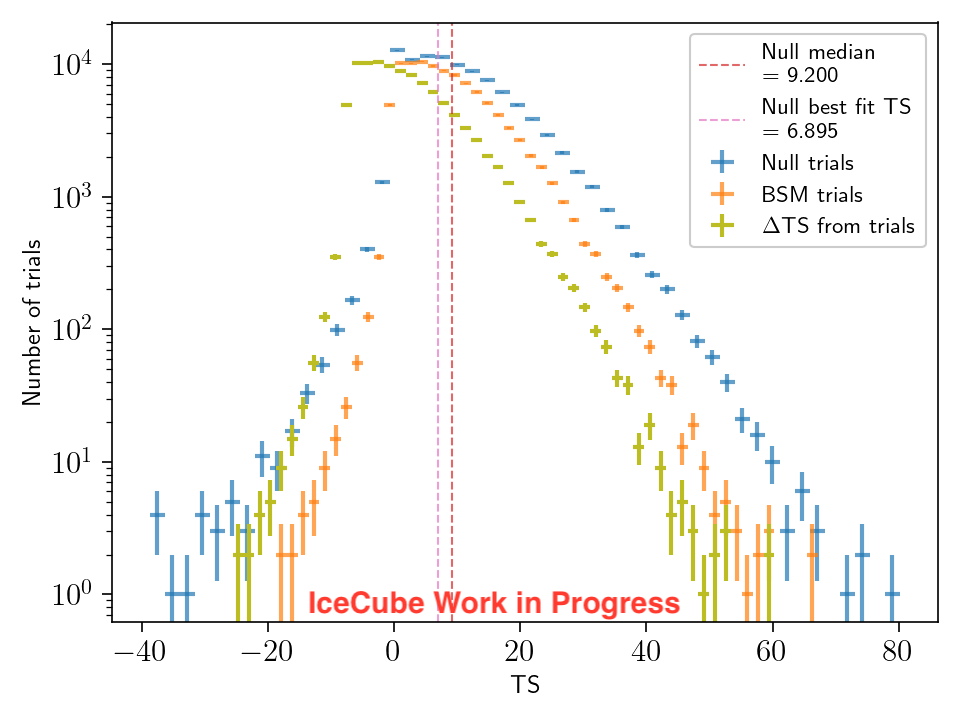}
    \includegraphics[width=0.329\linewidth]{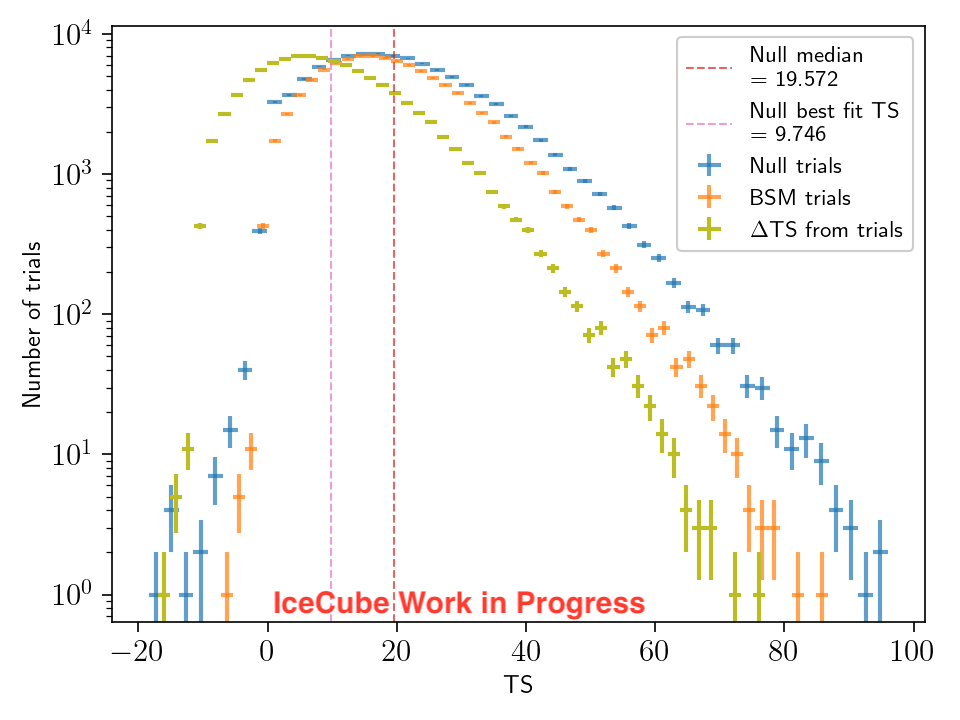}
    \caption{\textbf{Distributions of test statistics with the given numbers of the expected signal neutrinos} Test Statistic distributions from the likelihood ratio between the Background-only hypothesis and the Null hypothesis or the BSM hypothesis, in the blue and the orange histograms respectively, and $\Delta TS$ distribution in the green histogram. Each plot is derived by $10^5$ pseudo-experiments for each hypothesis with respect to the given numbers of expected signal neutrinos: $n_{s}=10$ (left), $n_{s}=25$ (middle), and $n_{s}=40$ (right).}
    \label{fig:TS_distributions}
\end{figure}

From a set of pseudo-experiments (trials) with given $n_{s}$, a set of log-likelihood values is obtained. Following the definition of $\mathcal{TS}$ in Equation~\ref{eq:test_statistics2}, the distributions of test statistics from the likelihood ratio between the Background-only hypothesis and the Null hypothesis or the BSM hypothesis can be derived, and $\Delta TS$ distribution can be secured as a consequence. $TS$ distributions and $\Delta TS$ distribution (coloured histograms) from the benchmark spectra in Figure~\ref{fig:benchmark_spectra} are shown in Figure~\ref{fig:TS_distributions} with three different $n_{s}$. The number of trials to derive each distribution was $10^5$. As shown in the plots, each distribution is distinguishable from the other distribution even visually. The preliminary sensitivity to the benchmark models will come from the hypothesis test between the Null hypothesis and the BSM hypothesis comparing the Null (blue) and the $\Delta TS$ (green) distributions.

\section{Summary and Outlook}\label{sec5}
This is the first search for neutrino signals having undergone rare BSM interactions with the whole IceCube astrophysical neutrino data and the sources of astrophysical neutrinos identified by IceCube. Considering a benchmark model for $\nu$-DM interaction, the analysis shows a prospect to test the BSM hypothesis that assumes the attenuation of astrophysical neutrino flux to the Null hypothesis with no attenuation by presenting the distinguishable test statistics distributions between both hypotheses. From the comparison of TS distributions, the sensitivity of this analysis to the benchmark BSM model can be achieved.

For the status, just one benchmark model is discussed in this proceeding which leads to a single local sensitivity. To have the global sensitivity over various models of $\nu$-DM interactions, BSM spectra need to be generated from diverse combinations of DM properties and interaction mediators and be added into the analysis framework for further evaluation. Furthermore, the analysis will implement systematic uncertainties including the astrophysical flux model introduced to characterise 9.5~years of the astrophysical $\nu_{\mu}$ data observed by IceCube~\cite{IceCube:2021uhz} for the Null hypothesis. The sensitivities among the different Null hypothesis assumptions will be evaluated from the tests of those models. Moreover, the framework can be applied to search for other rare interaction channels such as neutrino self-interaction considering the interaction of astrophysical neutrinos with $C \nu B$.

As the analysis utilises neutrino spectra in broad energies and precise event reconstructions, future neutrino detectors would provide improved sensitivities to the interactions. The IceCube-Gen2~\cite{IceCube-Gen2:2023ovv} is a proposed high-energy extension of the IceCube detector instrumenting an eight times larger effective volume than IceCube. The enlarged volume would increase the number of observed events even at much higher energies as well as enhance the angular resolution for track-like events.

\bibliographystyle{ICRC}
\bibliography{references}

\providecommand{\href}[2]{#2}\begingroup\raggedright\begin{thebibliography}{10}

\bibitem{Aartsen:2016nxy}
{\bfseries IceCube} Collaboration, M.~G. Aartsen {\em et~al.}
  \href{http://dx.doi.org/10.1088/1748-0221/12/03/P03012}{{\em JINST}
  {\bfseries 12} no.~03, (2017) P03012}.

\bibitem{IceCube:2013low}
{\bfseries IceCube} Collaboration, M.~G. Aartsen {\em et~al.}
  \href{http://dx.doi.org/10.1126/science.1242856}{{\em Science} {\bfseries
  342} (2013) 1242856}.

\bibitem{IceCube:2018dnn}
{\textbf{IceCube, Fermi-LAT, MAGIC, AGILE, ASAS-SN, HAWC, H.E.S.S., INTEGRAL,
  Kanata, Kiso, Kapteyn, Liverpool Telescope, Subaru, Swift NuSTAR, VERITAS,
  and VLA/17B-403} Collaborations, M.G. Aartsen \textit{et al.}}
  \href{http://dx.doi.org/10.1126/science.aat1378}{{\em Science} {\bfseries
  361} no.~6398, (2018) eaat1378}.

\bibitem{IceCube:2018cha}
{\bfseries IceCube} Collaboration, M.~G. Aartsen {\em et~al.}
  \href{http://dx.doi.org/10.1126/science.aat2890}{{\em Science} {\bfseries
  361} no.~6398, (2018) 147--151}.

\bibitem{IceCube:2022der}
{\bfseries IceCube} Collaboration, R.~Abbasi {\em et~al.}
  \href{http://dx.doi.org/10.1126/science.abg3395}{{\em Science} {\bfseries
  378} no.~6619, (2022) 538--543}.

\bibitem{IceCube:2023gpp}
{\bfseries IceCube} Collaboration, R.~Abbasi {\em et~al.}
  \href{http://dx.doi.org/10.1126/science.adc9818}{{\em Science} {\bfseries
  380} no.~6652, (2023) 1338--1343}.

\bibitem{Cherry:2014xra}
J.~F. Cherry, A.~Friedland, and I.~M. Shoemaker.
  \href{https://arxiv.org/abs/1411.1071}{arXiv:1411.1071}.

\bibitem{Barenboim:2019tux}
G.~Barenboim, P.~B. Denton, and I.~M. Oldengott
  \href{http://dx.doi.org/10.1103/PhysRevD.99.083515}{{\em Phys. Rev. D}
  {\bfseries 99} no.~8, (2019) 083515}.

\bibitem{Berryman:2022hds}
J.~M. Berryman {\em et~al.}
  \href{http://dx.doi.org/10.1016/j.dark.2023.101267}{{\em Phys. Dark Universe}
  {\bfseries 42} (2023) 101267}.

\bibitem{Bustamante:2020mep}
M.~Bustamante, C.~Rosenstr\o{}m, S.~Shalgar, and I.~Tamborra
  \href{http://dx.doi.org/10.1103/PhysRevD.101.123024}{{\em Phys. Rev. D}
  {\bfseries 101} no.~12, (2020) 123024}.

\bibitem{Bustamante:2018mzu}
M.~Bustamante and S.~K. Agarwalla
  \href{http://dx.doi.org/10.1103/PhysRevLett.122.061103}{{\em Phys. Rev.
  Lett.} {\bfseries 122} no.~6, (2019) 061103}.

\bibitem{Arguelles:2017atb}
C.~A. Arg\"uelles, A.~Kheirandish, and A.~C. Vincent
  \href{http://dx.doi.org/10.1103/PhysRevLett.119.201801}{{\em Phys. Rev.
  Lett.} {\bfseries 119} no.~20, (2017) 201801}.

\bibitem{IceCube:2015hes}
{\bfseries IceCube} Collaboration
  \href{http://dx.doi.org/10.22323/1.236.1081}{{\em PoS} {\bfseries ICRC2015}
  (2016) 1081}.

\bibitem{IceCube:2022clp}
{\bfseries IceCube} Collaboration, R.~Abbasi {\em et~al.}
  \href{https://arxiv.org/abs/2205.12950}{arXiv:2205.12950}.

\bibitem{Choi:2019ixb}
K.-Y. Choi, J.~Kim, and C.~Rott
  \href{http://dx.doi.org/10.1103/PhysRevD.99.083018}{{\em Phys. Rev. D}
  {\bfseries 99} no.~8, (2019) 083018}.

\bibitem{IceCube:2021uhz}
{\bfseries IceCube} Collaboration, R.~Abbasi {\em et~al.}
  \href{http://dx.doi.org/10.3847/1538-4357/ac4d29}{{\em Astrophys. J.}
  {\bfseries 928} no.~1, (2022) 50}.

\bibitem{Navarro:1996gj}
J.~F. Navarro, C.~S. Frenk, and S.~D.~M. White
  \href{http://dx.doi.org/10.1086/304888}{{\em Astrophys. J.} {\bfseries 490}
  (1997) 493--508}.

\bibitem{Nesti:2013uwa}
F.~Nesti and P.~Salucci
  \href{http://dx.doi.org/10.1088/1475-7516/2013/07/016}{{\em JCAP} {\bfseries
  07} (2013) 016}.

\bibitem{Planck:2018vyg}
{\bfseries Planck} Collaboration, N.~Aghanim {\em et~al.}
  \href{http://dx.doi.org/10.1051/0004-6361/201833910}{{\em Astron. Astrophys.}
  {\bfseries 641} (2020) A6}.

\bibitem{Farzan:2014gza}
Y.~Farzan and S.~Palomares-Ruiz
  \href{http://dx.doi.org/10.1088/1475-7516/2014/06/014}{{\em JCAP} {\bfseries
  06} (2014) 014}.

\bibitem{Ferrer:2022kei}
F.~Ferrer, G.~Herrera, and A.~Ibarra
  \href{http://dx.doi.org/10.1088/1475-7516/2023/05/057}{{\em JCAP} {\bfseries
  05} (2023) 057}.

\bibitem{Pandey:2018wvh}
S.~Pandey, S.~Karmakar, and S.~Rakshit
  \href{http://dx.doi.org/10.1007/JHEP11(2021)215}{{\em JHEP} {\bfseries 01}
  (2019) 095}.

\bibitem{Pandey:2020pgf}
S.~Pandey, S.~Karmakar, and S.~Rakshit.
  \href{https://doi.org/10.31526/ACP.NDM-2020.11}{\emph{ACP} \textbf{NDM-2020}
  (2020) 11}.

\bibitem{IceCube:2018ndw}
{\bfseries IceCube} Collaboration, M.~G. Aartsen {\em et~al.}
  \href{http://dx.doi.org/10.1140/epjc/s10052-019-6680-0}{{\em Eur. Phys. J. C}
  {\bfseries 79} no.~3, (2019) 234}.

\bibitem{IceCube:2019cia}
{\bfseries IceCube} Collaboration, M.~G. Aartsen {\em et~al.}
  \href{http://dx.doi.org/10.1103/PhysRevLett.124.051103}{{\em Phys. Rev.
  Lett.} {\bfseries 124} no.~5, (2020) 051103}.

\bibitem{Braun:2008bg}
J.~Braun, J.~Dumm, F.~De~Palma, C.~Finley, A.~Karle, and T.~Montaruli
  \href{http://dx.doi.org/10.1016/j.astropartphys.2008.02.007}{{\em Astropart.
  Phys.} {\bfseries 29} (2008) 299--305}.

\bibitem{IceCube-Gen2:2023ovv}
{\bfseries IceCube-Gen2} Collaboration {\em PoS} {\bfseries ICRC2023} (these
  proceedings) 994.

\end{thebibliography}\endgroup

\clearpage

\section*{Full Author List: IceCube Collaboration}

\scriptsize
\noindent
R. Abbasi$^{17}$,
M. Ackermann$^{63}$,
J. Adams$^{18}$,
S. K. Agarwalla$^{40,\: 64}$,
J. A. Aguilar$^{12}$,
M. Ahlers$^{22}$,
J.M. Alameddine$^{23}$,
N. M. Amin$^{44}$,
K. Andeen$^{42}$,
G. Anton$^{26}$
% ...
\\
\\
% ...
$^{12}$ Universit{\'e} Libre de Bruxelles, Science Faculty CP230, B-1050 Brussels, Belgium \\
% ...
$^{17}$ Department of Physics, Loyola University Chicago, Chicago, IL 60660, USA \\
$^{18}$ Dept. of Physics and Astronomy, University of Canterbury, Private Bag 4800, Christchurch, New Zealand \\
% ...
$^{22}$ Niels Bohr Institute, University of Copenhagen, DK-2100 Copenhagen, Denmark \\
$^{23}$ Dept. of Physics, TU Dortmund University, D-44221 Dortmund, Germany \\
% ...
$^{26}$ Erlangen Centre for Astroparticle Physics, Friedrich-Alexander-Universit{\"a}t Erlangen-N{\"u}rnberg, D-91058 Erlangen, Germany \\
% ...
$^{40}$ Dept. of Physics and Wisconsin IceCube Particle Astrophysics Center, University of Wisconsin{\textendash}Madison, Madison, WI 53706, USA \\
% ...
$^{42}$ Department of Physics, Marquette University, Milwaukee, WI, 53201, USA \\
% ...
$^{44}$ Bartol Research Institute and Dept. of Physics and Astronomy, University of Delaware, Newark, DE 19716, USA \\
% ...
$^{63}$ Deutsches Elektronen-Synchrotron DESY, Platanenallee 6, 15738 Zeuthen, Germany  \\
$^{64}$ Institute of Physics, Sachivalaya Marg, Sainik School Post, Bhubaneswar 751005, India \\
% ...

\subsection*{Acknowledgements}

\noindent
The authors gratefully acknowledge the support from the following agencies and institutions:
USA {\textendash} U.S. National Science Foundation-Office of Polar Programs,
U.S. National Science Foundation-Physics Division,
U.S. National Science Foundation-EPSCoR,
Wisconsin Alumni Research Foundation,
Center for High Throughput Computing (CHTC) at the University of Wisconsin{\textendash}Madison,
Open Science Grid (OSG),
Advanced Cyberinfrastructure Coordination Ecosystem: Services {\&} Support (ACCESS),
Frontera computing project at the Texas Advanced Computing Center,
U.S. Department of Energy-National Energy Research Scientific Computing Center,
Particle astrophysics research computing center at the University of Maryland,
Institute for Cyber-Enabled Research at Michigan State University,
and Astroparticle physics computational facility at Marquette University;
Belgium {\textendash} Funds for Scientific Research (FRS-FNRS and FWO),
FWO Odysseus and Big Science programmes,
and Belgian Federal Science Policy Office (Belspo);
Germany {\textendash} Bundesministerium f{\"u}r Bildung und Forschung (BMBF),
Deutsche Forschungsgemeinschaft (DFG),
Helmholtz Alliance for Astroparticle Physics (HAP),
Initiative and Networking Fund of the Helmholtz Association,
Deutsches Elektronen Synchrotron (DESY),
and High Performance Computing cluster of the RWTH Aachen;
Sweden {\textendash} Swedish Research Council,
Swedish Polar Research Secretariat,
Swedish National Infrastructure for Computing (SNIC),
and Knut and Alice Wallenberg Foundation;
European Union {\textendash} EGI Advanced Computing for research;
Australia {\textendash} Australian Research Council;
Canada {\textendash} Natural Sciences and Engineering Research Council of Canada,
Calcul Qu{\'e}bec, Compute Ontario, Canada Foundation for Innovation, WestGrid, and Compute Canada;
Denmark {\textendash} Villum Fonden, Carlsberg Foundation, and European Commission;
New Zealand {\textendash} Marsden Fund;
Japan {\textendash} Japan Society for Promotion of Science (JSPS)
and Institute for Global Prominent Research (IGPR) of Chiba University;
Korea {\textendash} National Research Foundation of Korea (NRF);
Switzerland {\textendash} Swiss National Science Foundation (SNSF);
United Kingdom {\textendash} Department of Physics, University of Oxford.

\end{document}